\pdfoutput=1 
\documentclass[%
superscriptaddress,
bibnotes,
amsmath,amssymb,
prb,
floatfix,
longbibliography,
twocolumn,
]{revtex4-2}

\usepackage{float}
\usepackage{color}
\usepackage[usenames,dvipsnames,svgnames,table]{xcolor}
\usepackage[colorlinks=true,linkcolor=blue,urlcolor=blue,citecolor=blue]{hyperref}
\usepackage{graphicx}
\usepackage{dcolumn}
\usepackage{bm}
\usepackage{xcolor}
\usepackage{enumerate}
\usepackage{braket}
\usepackage[T1]{fontenc}
\usepackage{listings}
\usepackage{tabularx}
\usepackage{booktabs}

\usepackage[english]{babel}
\makeatletter
\def\bbl@set@language#1{%
  \edef\languagename{%
    \ifnum\escapechar=\expandafter`\string#1\@empty
    \else\string#1\@empty\fi}%
  \@ifundefined{babel@language@alias@\languagename}{}{%
    \edef\languagename{\@nameuse{babel@language@alias@\languagename}}%
  }%
  \select@language{\languagename}%
  \expandafter\ifx\csname date\languagename\endcsname\relax\else
    \if@filesw
      \protected@write\@auxout{}{\string\select@language{\languagename}}%
      \bbl@for\bbl@tempa\BabelContentsFiles{%
        \addtocontents{\bbl@tempa}{\xstring\select@language{\languagename}}}%
      \bbl@usehooks{write}{}%
    \fi
  \fi}
\newcommand{\DeclareLanguageAlias}[2]{%
  \global\@namedef{babel@language@alias@#1}{#2}%
}
\makeatother

\DeclareLanguageAlias{en}{english}

\newcommand{\br}{\bm{r}}
\newcommand{\bR}{\bm{R}}
\newcommand{\ubr}{\underline{\bm{r}}}
\newcommand{\ubR}{\underline{\bm{R}}}

\newcommand{\s}{_\mathrm{{\scriptscriptstyle S}}}
\newcommand{\h}{_\mathrm{{\scriptscriptstyle H}}}
\newcommand{\xc}{_\mathrm{{\scriptscriptstyle XC}}}

\begin{document}

\preprint{ml-dft-review/1.0}

\title{A Deep Dive into Machine Learning Density Functional Theory for Materials Science and Chemistry}

\author{L. Fiedler}
\email{l.fielder@hzdr.de}
\affiliation{Center for Advanced Systems Understanding (CASUS), D-02826 G\"orlitz, Germany}
\affiliation{Helmholtz-Zentrum Dresden-Rossendorf, D-01328 Dresden, Germany}

\author{K. Shah}
\email{k.shah@hzdr.de}
\affiliation{Center for Advanced Systems Understanding (CASUS), D-02826 G\"orlitz, Germany}
\affiliation{Helmholtz-Zentrum Dresden-Rossendorf, D-01328 Dresden, Germany}

\author{M. Bussmann}
\email{m.bussmann@hzdr.de}
\affiliation{Center for Advanced Systems Understanding (CASUS), D-02826 G\"orlitz, Germany}
\affiliation{Helmholtz-Zentrum Dresden-Rossendorf, D-01328 Dresden, Germany}

\author{A. Cangi}
\email{a.cangi@hzdr.de}
\affiliation{Center for Advanced Systems Understanding (CASUS), D-02826 G\"orlitz, Germany}
\affiliation{Helmholtz-Zentrum Dresden-Rossendorf, D-01328 Dresden, Germany}

\date{\today}

\begin{abstract}
With the growth of computational resources, the scope of electronic structure simulations has increased greatly. Artificial intelligence and robust data analysis hold the promise to accelerate large-scale simulations and their analysis to hitherto unattainable scales. Machine learning is a rapidly growing field for the processing of such complex datasets. It has recently gained traction in the domain of electronic structure simulations, where density functional theory takes the prominent role of the most widely used electronic structure method. Thus, DFT calculations represent one of the largest loads on academic high-performance computing systems across the world.
Accelerating these with machine learning can reduce the resources required and enables simulations of larger systems.
Hence, the combination of density functional theory and machine learning has the potential to rapidly advance electronic structure applications such as in-silico materials discovery and the search for new chemical reaction pathways. We provide the theoretical background of both density functional theory and machine learning on a generally accessible level. This serves as the basis of our comprehensive review including research articles up to December 2020 in chemistry and materials science that employ machine-learning techniques. In our analysis, we categorize the body of research into main threads and extract impactful results. We conclude our review with an outlook on exciting research directions in terms of a citation analysis. 
\end{abstract}

\maketitle

\tableofcontents

\section{Introduction}
Electronic structure theory calculations enable the understanding of matter on the quantum level and complement experimental studies both in material science and chemistry. They play a central role in solving pressing scientific and technological problems. Large-scale electronic structure simulations have in turn been enabled by the advent of modern, high-performance computational resources. Yet, the ever increasing demand for accurate first-principles data renders even the most efficient simulation codes infeasible. On the other hand, the use of data-driven machine-learning methods has grown rapidly across a number of research fields. Such methods are increasingly gaining importance, as they are utilized to accelerate, replace, or improve traditional electronic structure theory methods.

\begin{figure}[h]
    \includegraphics[width=0.9\columnwidth]{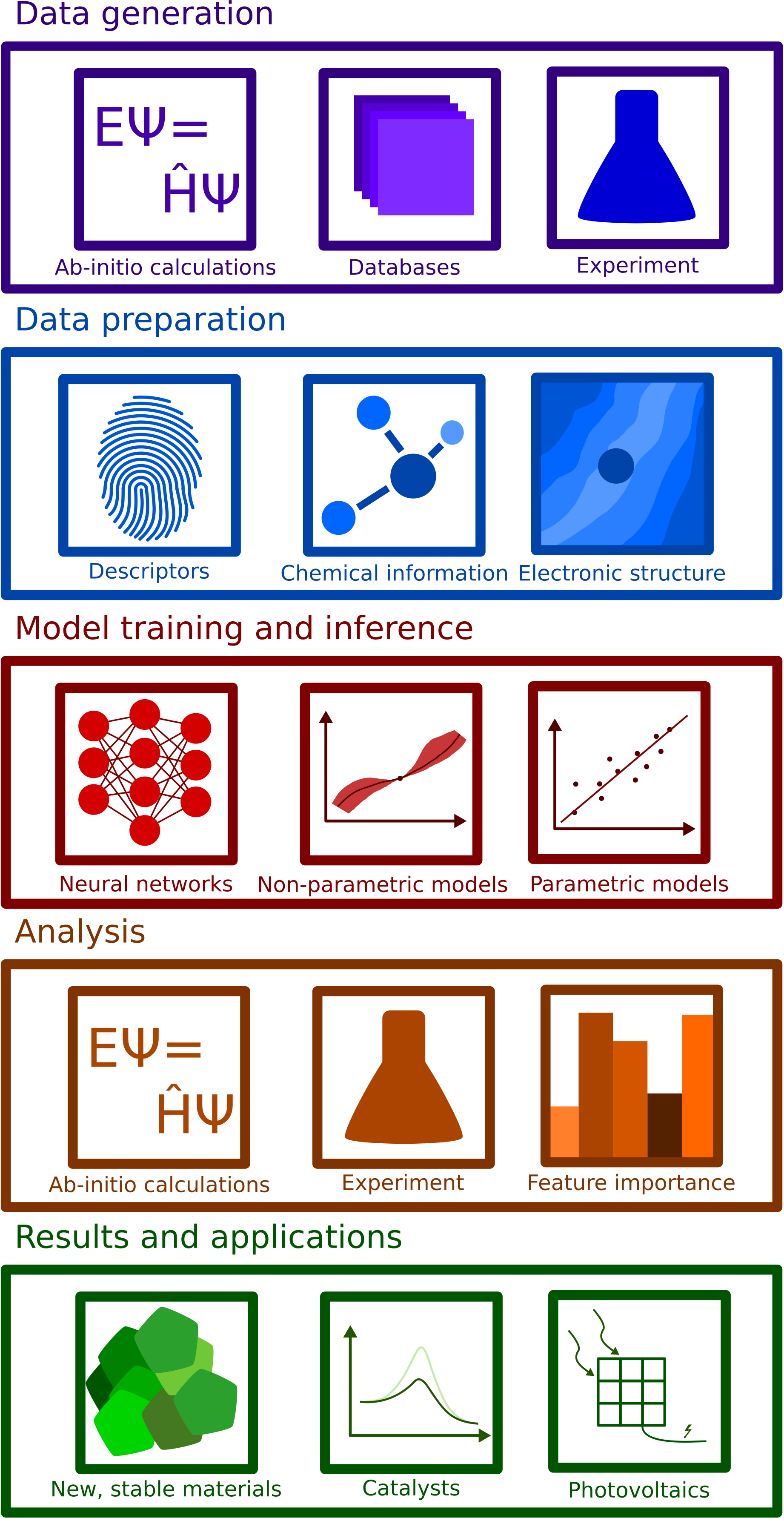}
    \label{fig:ml_workflow_outline}
    \caption{Schematic overview of the general workflow combining ML and DFT for material science or chemistry applications.}
\end{figure}

The computational backbone of electronic structure workflows is often density functional theory (DFT). While DFT provides a convenient balance between computational cost and accuracy, enormous speed-ups can be achieved when combined with machine learning (ML). In the following, we provide an overview over recent research efforts that accelerate materials science with the aid of ML and tackle different aspects of the combined ML-DFT workflow outlined schematically in Fig.~\ref{fig:ml_workflow_outline}. To that end, we compiled a database of over 300 research articles including the most recent publications up to December 2020. In Sec.~\ref{sec:theoretical_background} we introduce the necessary formalism and methodologies. Sec.~\ref{sec:review_section} is the centerpiece of our manuscript, where we provide an in-depth review of the compiled publications according to the most relevant research lines and highlight the most prominent applications. Finally, in Sec.~\ref{sec:discussion}, we carry out a citation analysis and point to trends of future research.

\section{Theoretical Background}
\label{sec:theoretical_background}

\subsection{Electronic Structure Theory}
\label{sec:electronic_structure}
Developing practical methods for calculating the electronic structure of materials is a topic of active research.
In the following, we provide a short overview of the basic formalism with a focus on DFT. We work within atomic units throughout, where $\hbar = m_e = e^2 = 1$, such that energies are expressed in Hartrees and length in Bohr radii.

\subsubsection{Exact framework} 
The most common theoretical framework for describing thermodynamic materials properties from first principles is within the scope of non-relativistic quantum mechanics.
It provides a systematic representation of $N_e$ electrons with collective coordinates $\ubr=\{\br_1,\dots,\br_{N_e}\}$ that are coupled to $N_i$ ions with collective coordinates $\ubR=\{\bR_1,\dots,\bR_{N_i}\}$, where $\br_j \in \mathbb{R}^3$ refers to the position of the $j$th electron, while $\bR_\alpha \in \mathbb{R}^3$ denotes the position of the $\alpha$th ion. Note that for sake of brevity we do not take spins into account.
The physics in this framework is governed by the Hamiltonian \cite{abedi_correlated_2012}
\begin{align}
    \label{eq.bo-hamiltonian}
    \hat{H}^{BO}(\ubr; \ubR) &= \hat{T}^{e}(\ubr) + \hat{V}^{ee}(\ubr) + \hat{V}^{ei}(\ubr; \ubR)+ E^{ii}(\ubR) \; ,
\end{align}
where $\hat{T}^e$ refers to the kinetic energy of the electrons, while the Coulomb operators, $\hat{V}^{ei}$ and $\hat{V}^{ee}$ account for the electron-ion and electron-electron interactions. In this equation, the Born-Oppenheimer approximation \cite{born_zur_1927} is employed to reduce the computational complexity of the underlying problem, i.e., the coupled electron-ion problem is separated and the ion-ion interaction amounts to a constant shift in energy $E^{ii}$. This approximation is valid as long as the motion of the ions happens on a much larger time scale than the motion of the electrons. Employing this Hamiltonian in the Schr\"odinger equation
\begin{equation}
	\hat{H}^{BO}(\ubr; \ubR) \Psi(\ubr; \ubR) = E \Psi(\ubr; \ubR) \label{eq:Schrödinger_stationary} \; ,
\end{equation}
with the many-electron wavefunction $\Psi$, which depends on the electronic coordinates and only parametrically on the ionic positions, provides the basis for the electronic structure of matter. It therefore enables quantitative predictions of physical phenomena in materials science. Its solution yields the electronic ground state based on which a vast amount of materials properties can be computed. Most importantly, the electronic ground-state energy can be identified as
\begin{equation}
\label{BO_PES}
    E = \bra{\Psi}\hat{H}^{BO}\ket{\Psi} \; .
\end{equation}
Other important quantities include molecular or crystal structures \cite{seth_crystal_2010}, charge densities \cite{fang_influence_2019}, cohesive energies \cite{civalleri_ab_2007}, elastic properties \cite{duan_phase_2015}, vibrational properties \cite{ben_yahia_updated_2009}, magnetic properties \cite{eustace_probing_2010, pantazis_new_2009}, dielectric susceptibilities \cite{andriyevsky_dft-based_2013}, magnetic susceptibilities \cite{mauri_magnetic_1996}, phase transitions \cite{la_porta_dft_2014}, bond dissociations \cite{khrapkovskii_formation_2010}, enthalpies of formation \cite{khrapkovskii_formation_2010}, ionization potentials \cite{zhan_ionization_2003}, electron affinities \cite{zhan_ionization_2003}, band gaps \cite{hernandez-haro_dft_2019, barhoumi_bandgap_2021}, and the equation of state \cite{daghash_structural_2019}. 

\subsubsection{Practical methods}
The multitude of practical electronic structure methods 
employ different approximations to overcome the underlying complexity of the problem. In the following concise overview of these methods, we always work within the Born-Oppenheimer approximation and suppress denoting the parametric dependence on $\ubR$.

\subparagraph{Density Functional Theory} 
\label{sec:dft}
DFT is a very popular method for carrying out electronic structure calculations, as it balances acceptable accuracy with reasonable computational cost.
The central quantity is the electronic density $n(\bm{r})$. The Hohenberg-Kohn theorems \cite{hohenberg_inhomogeneous_1964} guarantee a one-to-one correspondence between the electronic density and the external potential, e.g., the electron-ion potential $\hat{V}^{ei}(\ubr)$. This means every property of interest can be determined as a functional of the density. 

Practical DFT calculations usually rely on the Kohn-Sham ansatz \cite{kohn_self-consistent_1965} (KS-DFT). Here, the system of interacting electrons is replaced by an auxiliary system of non-interacting electrons constrained to reproduce the electronic density of the interacting system. This is achieved in terms of the Kohn-Sham equations
\begin{align}
	\left[-\frac{1}{2}\nabla^2 + v\s(\br)\right]\phi_j(\br) &= \epsilon_j \phi_j(\br) \; , \label{eq:KSequation.zerotemp}
\end{align}
a set of $N_e$ one-particle Schr\"odinger equations, where $v^\tau\s(\br)$ denotes the Kohn-Sham potential. The self-consistently calculated Kohn-Sham potential includes both the electron-ion interaction and the electron-electron interaction, the latter in terms of a mean-field description. The electronic density is obtained from the Kohn-Sham orbitals $\phi_j(\br)$ as
\begin{align}
    \label{eq:density.dft0K}
    n(\br) &= \sum_j \, |\phi_j(\br)|^2 \; .
\end{align}

The total energy functional is expressed within the Kohn-Sham formalism as
\begin{equation}
    E_\mathrm{total} = T\s[n] + E\h[n] + E\xc[n] + E^{ei}[n] + E^{ii} \label{eq:etotzerotemp}\; .
\end{equation}
In Eq.~(\ref{eq:etotzerotemp}), $E\h[n]$ denotes the Hartree energy, i.e., the energy contribution from the electrostatic interaction of the density with itself, $E^{ii}$ the energy contribution from the ion-ion interaction, and $E^{ei}[n]$ the energy contribution from the electron-ion interaction. $T\s[n]$ represents the kinetic energy of the Kohn-Sham system. The final energetic contributions are the exchange and correlation energies $E\xc[n]$. Given that the form of $E\xc[n]$ was known, it is ensured that DFT calculations yield \emph{in principle} the exact result. Note that the variational principle ensures that the total energy defined in Eq.~\ref{eq:etotzerotemp} remains stationary with respect to small changes in the density $n(\br)$. This leads to the definition of the Kohn-Sham potential in Eq.~\ref{eq:KSequation.zerotemp}.
In practice however, $E\xc[n]$ has to be approximated. The fundamental approximation to $E\xc[n]$ is the local density approximation (LDA)~\cite{kohn_self-consistent_1965, ceperley_ground_1980}, which by incorporation of the density gradient or further constraints can be expanded to (meta/hybrid-) generalized gradient approximations ((meta/hybrid-)GGA, e.g., PBE \cite{perdew_generalized_1996}, SCAN~\cite{sun_strongly_2015}, or B3LYP~\cite{becke_new_1993, lee_development_1988}). Orbital-free DFT (OF-DFT) \cite{karasiev_progress_2014,ligneres_introduction_2005,wesolowski_recent_2012,chen_orbital-free_2008} is an alternative to KS-DFT; it requires less computational power then the latter by \textit{purely} relying on density functionals to calculate energy terms, but is dependent on accurate approximations of the kinetic energy functional. Notable extensions of KS-DFT are time-dependent DFT (TD-DFT) \cite{runge_density-functional_1984} (representing the time-dependent Schr\"odinger equation in terms of the single-particle picture) and finite-temperature DFT (FT-DFT), which extends DFT to temperatures $\tau>0$ \cite{mermin_thermal_1965, graziani_thermal_2014, pittalis_exact_2011, karasiev_nonempirical_2013}.

\subparagraph{Beyond DFT} 
In contrast to DFT, wavefunction methods center on the many-particle wavefunction $\Psi(\ubr)$. The simplest methods is Hartree-Fock which constructs a single Slater determinant, but neglects electronic correlation by construction \cite{oliphant_systematic_1994}. Thus, several post-Hartree-Fock methods have been developed that incorporate electronic correlation explicitly, such as the coupled-cluster method (CC / CCSD / CCST) \cite{dunning_jr_gaussian_1989, woon_gaussian_1994, woon_gaussian_1993, koput_ab_2002, balabanov_systematically_2005, wilson_gaussian_1999}, configuration interaction (CI) \cite{siegbahn_generalizations_1980, lischka_new_1981, liu_alchemy_1981, saxe_shape-driven_1982, saunders_direct_1983, werner_self-consistent_1982, werner_matrix-formulated_1987, werner_comparison_1990, siegbahn_new_1984, knowles_chem_1984, dunning_gaussian_1977}, and M\o ller-Plesset perturbation theory (MP) \cite{moller_note_1934}. These methods generally outperform DFT calculations in terms of accuracy \cite{suellen_cross-comparisons_2019}, but do so at a much higher computational cost.

Exact electronic structure calculations can also be carried out using the density matrix renormalization group~\cite{DMRG} and Monte-Carlo methods such as variational Monte Carlo~\cite{wagner_qwalk_2009}, fixed-node diffusion Monte Carlo~\cite{kim_qmcpack_2018}, and path-integral Monte Carlo~\cite{barker_quantumstatistical_1979, militzer_first-principles_2021}.

\subparagraph{Coupling DFT to Molecular Dynamics} 
\label{sec:iap_intro}
Investigating the electronic structure of a material is often only a part of a bigger systematic study. Simulations on larger time and length scales are often realized by combining classical mechanical molecular dynamics (MD) of the ions with electronic structure calculations. To this end, one calculates an interatomic potential (IAP), which determines how the ions in the system interact. Such IAPs can also be replaced with reasonably fast quantum mechanical calculations (DFT-MD) or be constructed using electronic structure data for later use. Traditional IAPs are the Lennard-Jones potential \cite{jones_determination_1924}, the embedded atom model (EAM) \cite{daw_embedded-atom_1984} or other many-body potentials~\cite{daw_semiempirical_1983, daw_embedded-atom_1984, baskes_application_1987, tersoff_new_1988, brenner_empirical_1990, pettifor_analytic_1999, stuart_reactive_2000, van_duin_reaxff_2001, yu_charge_2007}. 
Recently, ML has given rise to a new class of IAPs that are constructed using data-driven algorithms (see Sec.~\ref{sec:ml}). Notable examples include ANI-1 \cite{smith_ani-1_2017}, GAP \cite{bartok_gaussian_2010}, BLAST \cite{chan_blast_2020}, HIP-NN \cite{lubbers_hierarchical_2018}, SchNet \cite{schutt_schnet_2018}, SNAP \cite{thompson_spectral_2015}, DPMD \cite{zhang_deep_2018}, and the AGNI ML force fields \cite{huan_universal_2017}. The resulting IAPs come well within chemical accuracy while retaining an often negligible computational cost.

\subsection{Machine learning}
\label{sec:ml}
ML is a form of computational pattern recognition which has had profound impact on multiple fields over the past decade. 
Advances in GPU hardware coupled with the development of high-level frameworks has led to many different applications in various fields. These applications include autonomous driving \cite{prakash_multi-modal_2021}, high-energy particle-physics data analysis \cite{schwartz_modern_2021}, and materials discovery \cite{pilania_machine_2021}. In this section, a brief background on ML is given, as well as some specifics relevant to most common models in materials science applications.

\subsubsection{Introduction}~
While there is no standard definition of ML, it is commonly defined as the study of computer algorithms that improve automatically through experience \cite{mitchell_machine_1997}. One usual way to divide ML models is into supervised models (models that make a prediction with respect to a ground truth with labeled data), unsupervised models (models that operate on data without an underlying ground truth with unlabeled data, and perform, e.g., dimensionality reduction) and reinforcement learning (models that act in complex environments to maximize rewards). ML models can be further divided into parametric and non-parametric models. When the model involves a finite set of learning parameters $\mathbf{W}$, the model is said to be parametric. These parameters are not set explicitly and learned iteratively through the learning process. Models without a finite set of parameters are non-parametric. In non-parametric models, the set of weights $\mathbf{W}$ is generally infinite.

Some examples of ML models include
\emph{supervised parametric models} (linear regression, neural networks), \emph{unsupervised parametric models} (Gaussian mixture models, generative adversarial networks~\cite{goodfellow_generative_2014}), \emph{supervised non-parametric models} (Gaussian Process Regression (GPR) \cite{rasmussen_gaussian_2006},  Decision Trees (DT) \cite{rokach_data_2008}, Support Vector Machines (SVM)~\cite{cortes_support-vector_1995}, Variational Autoencoders~\cite{kingma_auto-encoding_2014}), and \emph{unsupervised non-parametric models} (k-means algorithm) \cite{macqueen_methods_1967}.

Learning models can also be combined into ensembles. Using multiple learning models reduces overfitting and improves generalisability. Examples of ensemble learning techniques are Random Forests (Regression) (RF(R)) \cite{breiman_random_2001} and Gradient Boosting Regression (GB(R)) \cite{friedman_greedy_2001}.

Supervised parametric learning problems generally follow a common structure. Given a dataset $D$ consisting of $N$ samples $\mathbf{X}$ and labels $\mathbf{y}$, a model $M:\mathbf{X} \mapsto \bm{y}$ with parameters $W$ is to be fitted such that the cost or loss function $J$ is minimised. For this fitting only a subset of the dataset is used (training data), and once the model is sufficiently trained, its performance is analyzed on a separate subset of data (testing data). If the model performance is satisfactory, it can be used to predict $\bm{y}$ on input data for which it is unknown (inference). Care has to be taken during training so as to avoid over- or underfitting a model, i.e., introducing a high error on unseen data or not capturing the complexity of the underlying function, respectively. The optimization of the model is often done by gradient descent algorithms \cite{ruder_overview_2017} in an iterative fashion. Lastly, one has to consider so called hyperparameters, i.e., parameters that describe the model design and are set before model optimization  (e.g., characterization of the training process, kernel or activation functions). Their choice is crucial for reasonable model performance.

\subsubsection{Supervised Learning Models}
A brief description of commonly used supervised learning models in materials science is given in the following.
  
\subparagraph{Linear regression} 
Linear regression \cite{yan_linear_2009} is one of the most used algorithms for regression tasks. Given a dataset ($\mathbf{X},\bm{y}$) consisting of labeled data where $(\bm{x}^{(i)}, {y}^{*(i)}),~ i=1,...N$ with $\bm{x}^{(i)} \in \mathbf{X}$ and ${y^*}^{(i)} \in \bm{y}$, regression is used to determine the relationship between dependent variables $y^{*(i)}$ and independent variables $ \bm{x}^{(i)} = \{{x^{(i)}_1}, {x^{(i)}_2}, {x^{(i)}_3}, ... , {x^{(i)}_d}\} $, where $d$ is the dimension of the data. Given an input $\bm{x}^{(i)}$, the prediction $y^{(i)}$ is defined as 
\begin{equation}
y^{(i)} =  \mathbf{W}^T\bm{x}^{(i)} \,, \label{eq:linreg}
\end{equation}
where $\mathbf{W}$ are the learnt parameters of the model,  determined by minimizing a least squares cost function. While there exists an analytical solution for the optimal value of parameters $\mathbf{W}^*$, the computational complexity for calculating $\mathbf{W}^*$ is $\mathcal{O}(N^3)$ which becomes intractable for large data sets with high dimensions. 
If the number and range of the parameters is ill-suited for the dataset, the model might reach a sub-optimal state due to overfitting or underfitting. To overcome this, a regularisation term is added to the cost function; depending on the choice of this cost function one arrives at methods such as the Least Absolute Shrinkage and Selection Operator (LASSO)~\cite{tibshirani_regression_1996} or ridge regression \cite{hoerl_ridge_1970}. 
Sure Independence Screening and Sparsifying Operators (SISSO) \cite{ouyang_sisso_2018} is an extension to LASSO with better performance for feature selection. Kernel Ridge Regression (KRR) \cite{vovk_kernel_2013} is a non-parametric realisation of ridge regression. Instead of learned parameters $\mathbf{W}$, a kernel function measuring the distance between different data points is used. Linear regression models are not well suited for classification problems. When classification is needed, logistic regression \cite{hilbe_logistic_2009} can be applied. Here, the input data is mapped to discrete labels by applying a sigmoid function on the linear regression model.

\subparagraph{Neural networks}
Neural networks (NNs) are a promising approach to regression due to the \emph{universal approximation theorem} \cite{hornik_approximation_1991} that shows that any function can be modeled by a sufficiently sophisticated NN. 
NNs have been widely applied to problems in different domains including image classification~\cite{dosovitskiy_image_2020} or biomedical engineering~\cite{julie_handbook_2021}. However, neural networks also have some challenges in determining their hyperparameters and training routines that avoid overfitting.  

\paragraph{Introduction:}
A neural network is essentially a set of nested linear regression functions with non-linear activation functions. The base unit of a neural network is a perceptron or artificial neuron~\cite{rosenblatt_perceptron_1957}. 
For data point $\bm{x}^{(i)}$, the perceptron is defined as
\begin{equation}
f^{(i)} = \sigma(\mathbf{W}\bm{x}^{(i)} + b)\,,
\end{equation}
where $\sigma$ is a non-linear activation function, $\mathbf{W}$ consists of the weights, and $b$ denotes the bias. When nested together in layers, the model is known as multilayer perceptron or NN. The output of each layer is input to the subsequent layer. The first layer, that accepts the input samples, is called the input layer, and the final layer is called the output layer. The layers in between are referred to as hidden layers. A layer is defined by the number of its neurons which is often denoted as width. A NN is defined by the number and structure of hidden layers. A network with a large number of hidden layers is called a deep neural network (DNN). This gives rise to the term deep learning \cite{lecun_deep_2015}.
The design of a NN depends on the learning task. To successfully employ NNs, one has to make choices as to their activation functions~\cite{nwankpa_activation_2018} (e.g., Tanh and ReLU~\cite{nair_rectified_nodate}), loss function and optimization scheme (standard gradient descent can be replaced by stochastic gradient descent \cite{bottou_large-scale_2010} or the ADAM algorithm~\cite{kingma_adam_2017}).
  
\paragraph{Architectures:}
The operations performed by the neurons can be configured for each layer. The following are some of the most common architectures:

\begin{itemize}
\item[]  \emph{Fully Connected Neural Network (FCN) / Feed-Forward Neural networks:} Each neuron in a layer of FCNs is connected to every neuron in the next layer~\cite{minsky_perceptrons_1987}. 

\item[] \emph{Convolutional Neural Network (CNN):} CNNs are useful for tasks with 2D or 3D data such as image classification~\cite{dai_coatnet_2021}. 
 
\item[] \emph{Graph Neural Network (GNN):} Instead of a vector input, GNNs~\cite{scarselli_graph_2009,battaglia_relational_2018} can be used with data points in the form of arbitrary graph structures, useful for tasks like many-body problems in cosmology~\cite{cranmer_discovering_2020}. 
 
\item[]  \emph{Recurrent Neural Network (RNN):} RNNs are used for tasks that involve sequential or temporal evolution such as language translation~\cite{sutskever_sequence_2014}.

\item[] \emph{Physics Informed Neural Network (PINN):} PINNs are used to solve partial differential equations. This is done by incorporating initial and boundary conditions in the loss term~\cite{raissi_physics-informed_2019}.
\end{itemize}

NNs are effective at modelling complex non-linear functions and can be generalised well to new data~\cite{guan_analysis_2020}. Shortcomings of NNs are that large training datasets are required for practically useful results, and that the black-box models lack explainability and interpretability~\cite{cheng_interpretability_2021}.

\subparagraph{Gaussian Process Regression}
Gaussian Processes (GPs)~\cite{rasmussen_gaussian_2006} or Gaussian Process Regression (GPR) is a powerful non-parametric supervised learning technique with robust uncertainty quantification. It can be thought of as a distribution over functions. By constraining the probability distribution at training points, a learned function can be drawn over the test points. Instead of learning any learning parameters $\mathbf{W}$, inference is done by constructing the appropriate covariance matrix and drawing functions from it. Such a covariance matrix is defined by using a kernel function $k$, the choice of which impacts the usefulness of the model. With appropriate kernels and under certain conditions, GPs are equivalent to other ML methods such as neural networks~\cite{lee_deep_2018} and Kernel Ridge Regression~\cite{rasmussen_gaussian_2006}. GPs provide in-built uncertainty quantification and are effective for small datasets. However, the inference step involves matrix inversion which scales as $\mathcal{O}(N^3)$ for explicit inversion. This makes GPs computationally expensive for larger datasets. Approximate inference to increase the scalability of GPs is an active area of research~\cite{kuss_approximate_2005} and tractability has been greatly increased for sparse matrices with approaches such as Subset of Regressors~\cite{candela_unifying_2005-1}.



\section{Machine-learning in Materials Science and Chemistry}
\label{sec:review_section}
Electronic structure methods (Sec.~\ref{sec:electronic_structure}) suffer from their unfavorable computational scaling. Large scale calculations involving more than a few thousand atoms can only be realized with huge computational effort and time \cite{nakata_large_2020,bowler_calculations_2010}. Data-driven workflows (Sec.~\ref{sec:ml}) offer speedups by augmenting or replacing standard DFT calculations. 

\begin{figure*}[htp]
    \centering
    \includegraphics[width=1.9\columnwidth]{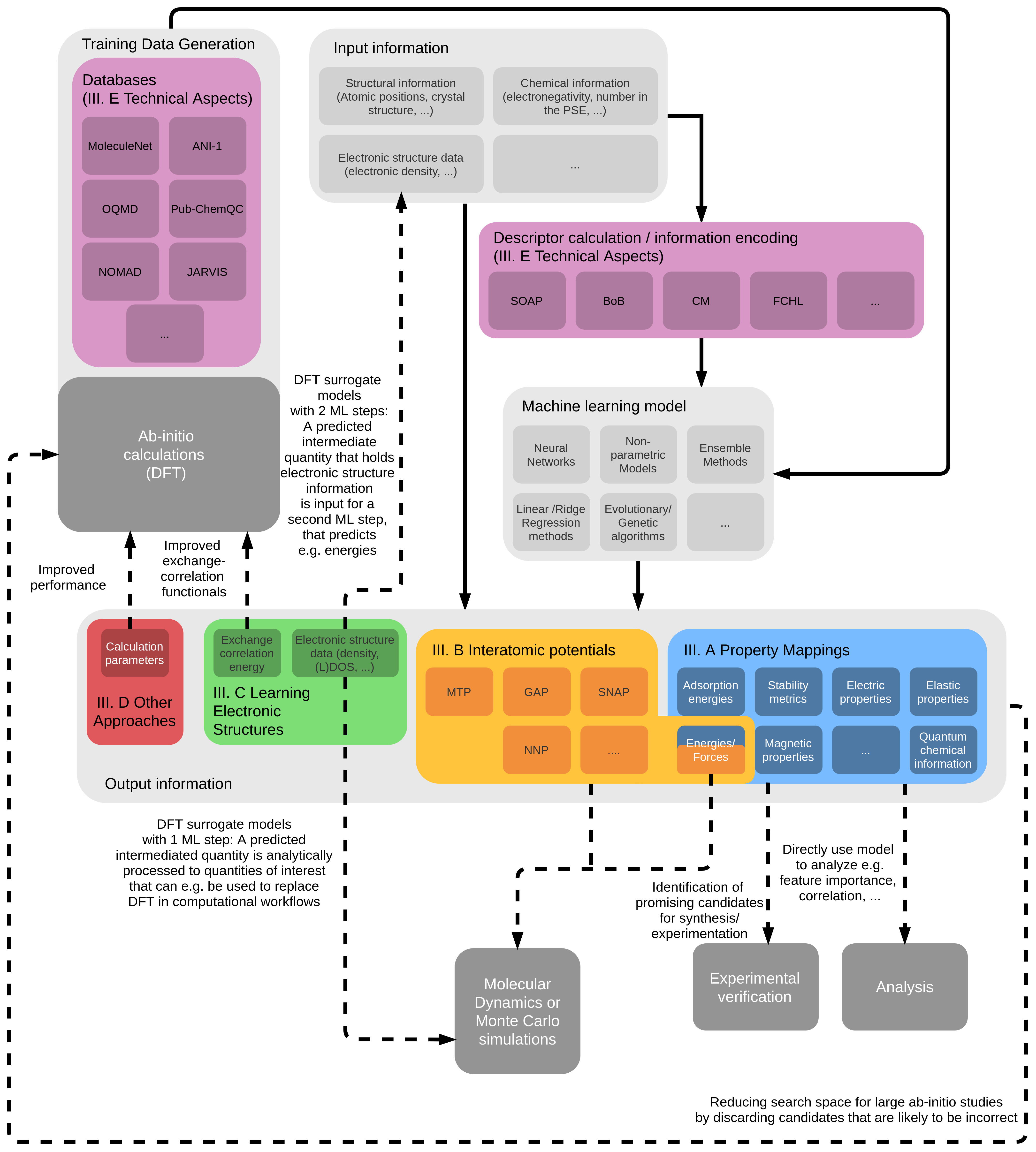}
    \caption{Color-coded map of workflows combining ML and DFT in current practice.}
    \label{fig:ml_workflow}
\end{figure*}

In the following, we provide an overview over the most recent research developments combining ML and DFT aimed at materials science and chemistry applications. 
We have grouped them into five main categories based on the type of approach they employ and we highlight the most representative articles. With this grouping, we identify the overall ML-DFT workflow visualized in Fig.~\ref{fig:ml_workflow}. By means of this diagram we link the individual categories presented in the following.

\subsection{Property mappings}
\label{sec:property_mappings}
Perhaps the most intuitive approach to employing ML techniques in DFT is by learning distinct physical or chemical properties from data sets. While accessing these through DFT calculations is computationally cheap when compared to other methods of comparable accuracy, the computational cost can often be restrictive when investigating large systems. Likewise, such investigations are often focused on a specific property relevant to a given application rather than on a fundamental understanding of the electronic structure. Hence a feasible approach is to train ML models to replicate a mapping of the form $\ubR \rightarrow p$ for a relevant property $p$ that one would initially obtain from a DFT calculation or from post-processing. In the following, an overview of such approaches is given, grouped by their respective application domain. Noteworthy high-impact results are discussed in detail.

\subsubsection{Chemical Reactions}
The demand to both reduce and capture carbon emissions motivates improving the efficiency of chemical reactions employed in industrial settings and searching for novel reactive pathways. In the context of computational materials science and chemistry this usually translates to identifying catalysts for reactions of interest (for instance $\mathrm{CO}_2$ reduction) and their respective activity and selectivity. The computational demand of such studies can be drastically decreased by employing ML, and a wide range of such investigations has been performed \cite{li_improving_2007, deng_understanding_2020, wu_rational_2020, fung_descriptors_2020, artrith_predicting_2020, liu_scalable_2020, praveen_design_2020, zheng_high-throughput_2020, ser_prediction_2020, friederich_machine_2020, ologunagba_machine_2020, rao_extendable_2020, lin_directly_2020, zhang_identifying_2020, sun_accelerating_2020, pedersen_high-entropy_2020, chowdhury_multiple_2020, kim_artificial_2020, lu_predicting_2020, saxena_silico_2020, li_improved_2019, guo_simultaneously_2019, back_convolutional_2019, hoyt_machine_2019, batchelor_high-entropy_2019, tahini_unraveling_2019, anderson_role_2018, toyao_toward_2018, wexler_chemical_2018, yada_machine_2018, jinnouchi_extrapolating_2017, ulissi_machine-learning_2017, gasper_adsorption_2017, boes_neural_2017, sawatlon_data_2019, li_feature_2017, meyer_machine_2018, wang_accelerating_2020, shetty_electric-field-assisted_2020, sun_machine-learning-accelerated_2020, li_computational_2020, maley_quantum-mechanical_2020, yang_machine_2020,hakala_hydrogen_2017, jinnouchi_predicting_2017, ulissi_address_2017, lu_molecular_2019, scherbela_charting_2018}.

A typical target quantity is the adsorption energy of a molecule on a catalyst. In Ref.~\cite{pedersen_high-entropy_2020}, the $\mathrm{CO}_2$ reduction is analyzed where its efficiency is highly dependent on the adsorption properties of CO and H on the chosen catalyst. The catalyst was a high entropy alloy (HEA) consisting of at least five elements \cite{yeh_nanostructured_2004}. The configuration space of possible potential surface adsorption sites is vast. A GPR model built on a subset of those enables the prediction of adsorption energies on the entire configuration space. In this manner, a candidate catalyst for this reaction was identified, which was then also studied experimentally \cite{nellaiappan_nobel_2019}. Similarly, Ref.~\cite{friederich_machine_2020} also uses GPR to predict activation barriers of dihydrogen on metal complexes. There are various approaches of using ML models to facilitate the design or discovery of catalysts by performing such high-throughput searches. Other notable examples include the $\mathrm{N}_2$ electroreduction on transition metal alloys using GNNs \cite{kim_artificial_2020} and the combination of linear models based on experimental data and RFR/GBR models on ab-initio data to directly predict activities and selectivities for ethanol reforming on bi-atom catalysts\cite{artrith_predicting_2020}.

Ref.~\cite{deng_understanding_2020}, which is concerned with the $\mathrm{O}_2$ reduction on bi-atom catalysts, follows a slightly different approach. Here, ML is used to unveil correlations in the underlying DFT data. By training a RF regressor to correctly predict adsorption energies from a number of chemical descriptors, the correlation between the adsorption energy and these descriptors is evaluated. Using ML to identify the relative importance of individual descriptors, i.e., catalyst design features, is a technique used for a variety of reaction/catalyst combinations. These include $\mathrm{CO}_2$ capture on MOFs \cite{anderson_role_2018}, $\mathrm{O}_2$ reduction on single atom catalysts \cite{guo_simultaneously_2019}, olefin oligomerization on Cr based catalysts \cite{maley_quantum-mechanical_2020}, and the hydrogen evolution reaction (HER) \cite{wexler_chemical_2018}.

\subsubsection{Discovery of Novel, Stable Materials}
\label{sec:stablematerials}
The discovery of novel materials is at the very core of computational materials science, as computational methods reduce the cost of exploring large configuration spaces drastically compared to experimental methods. The ever increasing complexity and size of these configuration spaces makes their exploration an ideal application for ML methods that can be employed to accelerate such investigations \cite{hu_accelerated_2020, zhang_machine_2020, yasin_machine_2020, siriwardane_revealing_2020, ulenberg_prediction_2015, meredig_combinatorial_2014, noh_uncertainty-quantified_2020, wu_global_2019, nandy_machine_2019, jennings_genetic_2019, li_self-evolving_2019, li_thermodynamic_2019, kim_machine-learning-accelerated_2018, zheng_machine_2018, ye_deep_2018, he_metallic_2018, li_predicting_2018, janet_accelerating_2018, chandran_machine_2018, seko_matrix-_2018, ward_including_2017, schmidt_predicting_2017, kaneko_regression_2019, hartnett_prediction_2019, park_developing_2020, xiong_combined_2020, schleder_exploring_2020, lu_predicting_2019, li_center-environment_2020, askerka_learning--templates_2019, jorgensen_machine_2018, curtis_gator_2018, nebgen_transferable_2018, choudhuri_local_2021}.

Such studies usually sample a certain space of possible candidate structures using some measure of stability, and possibly, specific values for properties of interest. Ref.~\cite{ye_deep_2018} uses DNNs and a minimal set of chemically motivated descriptors (electronegativity and ionic radii) to analyze a range of mixed-series crystals and perovskites. The formation energy is then predicted for these materials within chemical accuracy but at negligible computational cost. This enables rapid identification of stable crystal structures which can then be synthesized in the lab. A similar approach is employed in Ref.~\cite{xiong_combined_2020}, where a class of multicomponent Ti-alloys are investigated. It extends the previous study by prediciting elastic properties, which are important for aerospace applications. The combination of stability metrics with application-centered properties is also demonstrated in Ref.~\cite{schleder_exploring_2020}, where a SISSO model is trained to predict formation enthalpies of 2D materials to assess their thermodynamic stability. The proposed application domain of this method is the identification of materials suitable for photoelectrocatalytic water splitting. Here, the band gap is the primary metric. While the presented model is not capable of predicting the band gap, it significantly speeds up the search for structures by reducing the search space by more than a half. A similar example of screening a chemical for thermodynamic stability is Ref.~\cite{wu_global_2019}, where hybrid organic-inorganic perovskites (HOIP) for photovoltaics are at the center of attention.

\subsubsection{Electric Properties and Photovoltaics}
Driven by the effort towards a sustainable future, another important application domain is the search for materials useful in electronic applications such as batteries for electric cars, efficient photovoltaics, and semiconductors for microelectronics. While the underlying problem is similar to before, i.e., finding suitable materials from large configuration spaces, here the constraints on suitability are based on electrical properties such as the band gap or the photoelectric conversion coefficient (PCE). ML has been employed in a number of studies \cite{wu_deep_2020, meftahi_machine_2020, allam_molecular_2020, gomez-bombarelli_turbocharged_2016, gaultois_perspective_2016, mannodi-kanakkithodi_author_2020, janet_accurate_2020, zhu_fundamental_2020, choudhary_accelerated_2019, huang_band_2019, park_exploring_2019, padula_combining_2019, sendek_machine_2019, paul_transfer_2019, mannodi-kanakkithodi_machine_2019, jorgensen_machine_2018, rajan_machine-learning-assisted_2018, kurkova_discovering_2018, allam_application_2018, fernandez_machine_2017, zhang_two-dimensional_2021, ju_accelerated_2020, kauwe_extracting_2020, choudhary_data-driven_2020, frey_machine_2020, antono_machine-learning_2020, lourenco_adaptive_2020, choudhary_high-throughput_2020, schutt_how_2014}. 

A lot of studies are directed towards photovoltaic materials, for which the PCE, i.e., the percentage of light transferred to electrical energy by a photovoltaic material, is one of the most important metrics. In Ref.~\cite{meftahi_machine_2020}, Bayesian-Regularized NNs were trained on DFT data and used to predict the PCE for organic photovoltaic materials (OPVs). A prediction accuracy of $\pm 0.5\%$ 
was achieved, allowing for efficient screening of promising materials. The basis for these models are simple, chemical descriptors. 
A similar strategy has been employed in Ref.~\cite{padula_combining_2019}, where simple structural and electronic similarity features served as inputs for KRR models, again for OPVs. The investigation of photovoltaic materials is not limited to PCE models. 

Another important target quantity for ML approaches is the band gap, as demonstrated in Ref.~\cite{wu_deep_2020}, where the vast chemical space of HOIP is sampled, similar to the aforementioned Ref.~\cite{wu_global_2019}, however with an emphasis on the electrical properties of candidate materials. Using a small set chemical descriptors band gaps computed using DFT were learned and predicted by GBR. The results of Ref.~\cite{wu_global_2019} were incorporated in this investigation by means of progressive ML~\cite{fayek_progressive_2020}, in order to enrich the underlying database of the ML model. Yet, such models suffer from the underestimation of band gaps by usual DFT functionals, an issue addressed in Ref.~\cite{rajan_machine-learning-assisted_2018}. The established way of overcoming these inaccuracies inherent to DFT are by using many-body theories, such as the GW approximation, which are computationally very costly. By using GPR models, the corrections applied by the GW approximation~\cite{hedin_new_1965, aulbur_quasiparticle_2000, aryasetiawan_gw_1998} to band gaps obtained from DFT can be evaluated in a fraction of the time and independently from DFT, since the GPR models are only directly dependent on chemical descriptors. 

Other electrical properties of interest addressed in ML-based studies are piezoelectric and dielectric responses of inorganic materials \cite{choudhary_high-throughput_2020}, Li ion conductivity for Li ion batteries \cite{sendek_machine_2019} or impurity levels of semiconductors \cite{mannodi-kanakkithodi_author_2020}.

\subsubsection{Elastic and Structural Properties}
The design and discovery of materials with specified structural properties constitutes another important research thrust where DFT data and property mappings in terms of ML play a central role. These investigations \cite{salvador_discovery_2020, tong_accelerating_2020, rupp_machine_2015, jain_bulk_2019, tawfik_efficient_2019, musil_machine_2018, balachandran_predicting_2017, wang_new_2017, tamura_fast_2017, wang_error_2017, zhao_integration_2019, ayyasamy_density_2020} are similar in nature to the ones discussed in Sec.~\ref{sec:stablematerials}, however, with a stronger emphasis on property selection.

Typical quantities of interest in this context are elastic constants or Young's modulus. NNs and RFRs were used in Ref.~\cite{salvador_discovery_2020} to predict bulk, shear, and Young's modulus for ternary Ti-Nb-Zr alloys in order to accelerate an otherwise costly high-throughput DFT search for low-modulus alloys, similar to Ref.~\cite{xiong_combined_2020}. Identification of such alloys is in demand from the biomedical industry, and due to the ML assisted study, a promising candidate could be found. 

Another important property often considered in high-throughput calculations is the elastic constant. This is addressed by Ref.~\cite{wang_new_2017} and \cite{wang_error_2017} in terms of NN models to provide both direct predictions of elastic constants. 

ML models have been also used to unveil relationships between structure and properties of materials, such as in Ref.~\cite{musil_machine_2018} by means of a GPR based analysis of molecular crystals. Here, similarities between crystal structures were classified along with the prediction of lattice energies where the resulting structural landscapes shed light on molecular crystallization.

\subsubsection{Quantum Chemical Information}
\label{sec:quantumchemicalinformation}
While there is no precise distinction between quantum chemistry and materials science, these closely related fields can by distinguished in terms of either the systems considered or the properties under investigation. While the articles and groupings of property mappings discussed above are drawn from both quantum chemical and materials science applications, ML approaches are also heavily utilized for the rapid prediction of quantum chemical properties, such as polarizabilities, partial charges, HOMO-LUMO levels or bond dissociation energies \cite{balabin_neural_2009, liu_rapid_2020, bag_machine_2020, veit_predicting_2020, st_john_prediction_2020, gao_machine_2016, zhang_qspr_2014, qu_big_2013, li_accurate_2013, li_promising_2012, ramakrishnan_electronic_2015, shaikh_strain-tuned_2020, gugler_enumeration_2020, jha_enhancing_2019, wang_machine_2019, li_efficient_2019, wilkins_accurate_2019, iype_machine_2019, pereira_machine_2018, pronobis_capturing_2018, hy_predicting_2018, wang_significantly_2018, bleiziffer_machine_2018, faber_prediction_2017, janet_predicting_2017, pereira_machine_2017, li_machine_2020, liu_structure_2020, venkatraman_designing_2018, eckhoff_predicting_2020, duan_semi-supervised_2020, zhang_machine_2016, tsubaki_fast_2018}. 

In Ref.~\cite{wilkins_accurate_2019}, polarizabilities are calculated with both DFT and CCSD. The latter provides better overall accuracy at a drastically increased computational cost. GPR models trained on either of the resulting data sets are capable of predicting polarizabilities accurately. Yet the best performance with respect to the baseline provided by CCSD is achieved when learning and predicting the difference $\Delta$ between these two levels of theory and then applying the output of such a model to DFT results. Such an approach has been followed in combined ML/DFT applications (see Sec.~\ref{sec:learning_electronic_structure} and the approaches for exchange-correlation functionals discussed therein) and is often referred to as $\Delta$-learning \cite{ramakrishnan_big_2015}. It was originally introduced for the prediction of energies. Another example in the context of these target quantities is Ref.~\cite{gao_machine_2016}, where NNs predict corrections on top of DFT-level energies for non-covalent interactions. 

Other approaches follow an outline more consistent with the workflows described above, such as Ref.~\cite{st_john_prediction_2020}, wherein a GNN is utilized to predict bond dissociation energies for a large database of organic molecules. The resulting model can be applied for instance in drug design. Similarly, in  Ref.~\cite{janet_predicting_2017} NNs are used to predict spin-state splittings of transition metal complexes, employing a test data set that includes experimental data in addition to DFT data. 

Related to these works are applications of TDDFT, such as Ref.~\cite{ramakrishnan_electronic_2015}, wherein corrections to DFT calculated electronic spectra are learned from CCSD calculations within the $\Delta$-learning methodology. 

Finally, another interesting approach is presented in Refs.~\cite{liu_rapid_2020} and \cite{duan_semi-supervised_2020}. Models based on KRR and NN were employed to predict whether a specific configuration should be treated with DFT or more accurate (and costly) multireference  methods.

\subsubsection{Magnetic, Thermal, and Energetic Properties}
\label{sec:misc_properties}
Finally, there are ML property mappings that do not directly fall into the application-based groupings above. 
Properties that can be derived from DFT and predicted or analyzed with ML methods but have not yet been discussed at length are, for instance, transport and diffusion properties \cite{eremin_ionic_2019, wu_robust_2017, juneja_unraveling_2020}, surface energies \cite{palizhati_toward_2019}, and magnetic properties \cite{rhone_data-driven_2020}. 

Some studies that combine ML and DFT fall into scientific disciplines not mentioned yet, such as studying RNA conformations~\cite{icazatti_classification_2019} or the DFT prediction of nuclear magnetic resonance shifts (NMR) \cite{navarro-vazquez_dftmachine-learning_2020, chaker_nmr_2019, paruzzo_chemical_2018}. DFT can also be used as a pre-processing or pre-screening tool for ML workflows based on experimental data, as demonstrated in Ref.~\cite{balachandran_structure--curie_2016} and  Ref.~\cite{balachandran_data-driven_2020}, where Curie and transition temperatures are learned, respectively. A purist's approach to combining DFT and ML is taken in Ref.~\cite{seko_machine_2014}, which is concerned with melting temperatures. 

Lastly, an important direction of research is concerned with energetic materials, i.e., materials that hold large amounts of chemical energy, such as fuels or explosives. A comprehensive study of ML techniques for energetic materials can be found in Ref.~\cite{elton_applying_2018}.

\subsection{Interatomic Potentials}
\label{sec:iaps}
A specialized type of property mappings in terms of ML are interatomic potentials (IAPs) (Sec.~\ref{sec:iap_intro}). Here, a ML mapping of atomic positions to the potential energy and atomic forces is performed, also referred to as force field or potential energy surface (PES). Most commonly, these ML-IAPs parametrize the Born-Oppenheimer PES defined in Eq.~(\ref{BO_PES}). The trained models can be used to perform MD simulations at an accuracy close to the first-principles methods used for the calculation of the training data (e.g., DFT) with no significant computational overhead. In the following, an overview of developments in the field of ML-IAPs is given
The reviewed publications have been grouped by the type of IAP used. We note that many of these ML-IAPs are related to each other in terms of linear transformations or special cases of the atomic cluster\cite{lysogorskiy_performant_2021}.

\subsubsection{Neural Network Potentials}
A large number of IAP workflows rely on neural networks in order to approximate energies and forces from atomic positions \cite{morawietz_density-functional_nodate, babar_accurate_2020, hong_training_2020, gartner_signatures_2020, wu_modeling_2020, stricker_machine_2020, houchins_accurate_2020, lee_crystallization_2020, rodriguez_spatial_2020, rao_accelerated_2020, groenenboom_halide-induced_2020, gaillac_speeding_2020, hu_neural_2020, thorn_toward_2019, wen_development_2019, ko_isotope_2019, sosso_harnessing_2019, meshkov_sublattice_2019, galvelis_scalable_2019, smith_approaching_2019, eckhoff_molecular_2019, pun_physically_2019, huang_density_2019, kang_first-principles_2018, tsubaki_fast_2018, rostami_optimized_2018, liu_constructing_2018, li_study_2017, gao_modeling_2018, wang_crystal_2020, liang_molecular_2020, wang_molecular_2020, nagai_self-learning_2020, sosso_neural_2012, gabardi_atomistic_2017, pan_dft_2020, pan_dft_2021, korolev_accelerated_2020, nigussa_application_2020}.

The type, architecture, and data representation of the employed NNs vary drastically between different approaches. Arguably the most common type of NNs within these approaches are deep feed-forward networks as used in Ref.~\cite{gartner_signatures_2020}. Using such \emph{deep potentials}, the authors were able to investigate a proposed liquid-liquid state transition in water, a study unattainable with regular DFT calculations. Similarly, in Ref.~\cite{sosso_harnessing_2019} an investigation of the phase-change material GeTe, which is highly relevant for the development of non-volatile memory devices, is presented. These simulations would have been challenging with classical IAPs. Instead, a NN potential is used, drawing on Ref.~\cite{sosso_neural_2012} and Ref.~\cite{gabardi_atomistic_2017}. The transferability of models highlights another advantage of ML-IAPs in comparison to highly specialized workflows, as they aim to replace DFT or classical potentials with often no application-induced constraints.

Traditional NNs can be improved by incorporating physical considerations, such as in Ref.~\cite{pun_physically_2019}. The thought behind developing such physically enhanced networks is to enable extrapolation capabilities outside of the range of training data by encoding information on the chemical bonding of the atomic structures. Extrapolation is traditionally a problematic task for NNs, yet often necessary when performing simulations on new materials. They demonstrate the improved capabilities of such IAPs based on physically enhanced NNs for the case of aluminum.

The combination of DFT and ML is also leveraged to construct IAPs with an accuracy better than DFT itself, as shown in Ref.~\cite{smith_approaching_2019}. There, transfer learning is used to construct an IAP with CCSD accuracy. First an IAP is calculated using DFT data which is retrained via transfer learning on CCSD data. This enables rapid predictions of total energies with CCSD accuracy. A similar approach is taken in Ref.~\cite{pattnaik_machine_2020}. Here, a neural network is trained to correct a classical IAP to within DFT accuracy by training it on DFT data. Both of these approaches follow the principal idea of $\Delta$-learning mentioned in Sec.~\ref{sec:quantumchemicalinformation}.

\subsubsection{Gaussian Approximation Potentials}
Another important class of IAPs are Gaussian Approximation Potentials (GAPs), primarily based on Ref.~\cite{bartok_gaussian_2010}. Based upon GPs, the total energy is decomposed into atomic contributions, which are then learned by GPR from DFT data. Information on the atomic neighborhoods is encoded by suitable descriptors, e.g., based on a bispectrum representation of the atomic density (such as the SOAP descriptor, see \ref{sec:technical_aspects}). A range of publications have employed GAPs to model the PES \cite{rowe_development_2018, bartok_machine-learning_2013, konstantinou_simulation_2020, tovey_dft_2020, deringer_general-purpose_2020, liu_machine_2020, thiemann_machine_2020, rowe_accurate_2020, gillan_first-principles_2013, zhang_gaussian_2019, tong_accelerating_2018, mocanu_modeling_2018, deringer_towards_2018, dragoni_achieving_2018, deringer_extracting_2017, deringer_machine_2017, huang_first-principles_2019, bernstein_novo_2019, sosso_harnessing_2019}.

For example, in Ref.~\cite{deringer_machine_2017}, a GAP is trained for amorphous carbon. This potential is then used by the same authors in Ref.~\cite{deringer_towards_2018} to investigate electrode materials based on amorphous carbon through a combination of GAP and MD, a task usually unattainable with quantum mechanical accuracy. The same GAP is also employed in Ref.~\cite{huang_first-principles_2019} and \cite{deringer_extracting_2017}, further highlighting the universal re-usability of these IAPs.

A GAP has also been used to model phase-change materials~\cite{mocanu_modeling_2018}, similar to Ref.~\cite{sosso_harnessing_2019} which was based on NNs. As one would expect, there exists no singular optimal choice for IAPs, and several different ML workflows can be used to tackle similar problems. 
Furthermore, GAPs have been shown to be able to model complicated PES of magnetic materials in Ref.~\cite{dragoni_achieving_2018}.

Finally, GAPs can be incorporated in larger ML workflows, such as the CALYPSO structure prediction method~\cite{tong_accelerating_2018}. They also facilitate automatic workflows for IAP construction as given in Ref.~\cite{bernstein_novo_2019} or the FLARE library \cite{vandermause_--fly_2020}.

\subsubsection{Moment Tensor Potentials}
Similar to GAPs, moment tensor potentials (MTPs)~\cite{shapeev_moment_2016} are based on linear regression and polynomial expansion. They can, in principle, approximate any PES. MTPs pose another class of powerful ML-IAPs that have been in used in many publications \cite{mortazavi_accelerating_2021, mortazavi_machine-learning_2020, podryabinkin_accelerating_2019, wang_ionic_2020, podryabinkin_active_2017, gubaev_machine_2018, grabowski_ab_2019, jafary-zadeh_applying_2019, ladygin_lattice_2020, novoselov_moment_2019, gubaev_accelerating_2019, korotaev_accessing_2019, wang_lithium_2020, novikov_improving_2019, novikov_ring_2019, novikov_automated_2018, mortazavi_exploring_2020, mortazavi_efficient_2020, mortazavi_nanoporous_2020, raeisi_high_2020}, often in conjunction with active learning approaches. 

For example, Ref.~\cite{wang_ionic_2020} investigates solid-state batteries, which are of high interest due to the increasing demand for powerful and reliable electrical power sources. They employed a MTP that is trained on the fly that improves the accuracy of the IAP with ab-initio data where needed. 

Using MTPs for active learning was first proposed by the authors of Ref.~\cite{shapeev_moment_2016}. A linearly parameterized IAP is constructed on the fly based on the D-optimality criterion~\cite{podryabinkin_active_2017}. This method has subsequently been used to study diffusion processes~\cite{novoselov_moment_2019} or lattice dynamics~\cite{ladygin_lattice_2020}. As the authors point out, this technique is not limited to MTPs and can be employed for other linear IAPs as well.

Another prominent application of MTPs is the investigation of both phononic properties and thermal conductivities~\cite{mortazavi_accelerating_2021,mortazavi_exploring_2020,mortazavi_machine-learning_2020,raeisi_high_2020} and the search for new alloys \cite{gubaev_accelerating_2019}.

\subsubsection{Spectral Neighborhood Analysis and Other Potentials}
Given the amount of possibilities ML-DFT techniques provide, it is no surprise that there are a number of IAP construction methods~\cite{zeni_building_2018, chen_accurate_2017, nishiyama_application_2020, pan_dft_2020, chapman_multiscale_2020, nigussa_application_2020, korolev_accelerated_2020, pihlajamaki_monte_2020, rupp_machine_2014, seko_first-principles_2015, bernstein_novo_2019, li_quantum-accurate_2018, scherbela_charting_2018, narayanan_machine_2017, miwa_molecular_2016, pan_dft_2021, thompson_spectral_2015, wood_data-driven_2019} that do not fall into the three categories outlined above.

The most prominent of these is the spectral neighborhood analysis potential (SNAP)~ \cite{thompson_spectral_2015, cusentino_explicit_2020, wood_extending_2018, wood_data-driven_2019}.
It is based on a linear model, similar in construction to GAP and MTPs. 
SNAP has been used to investigate lithium nitride~\cite{deng_electrostatic_2019}, carbon under extreme conditions~\cite{willman_quantum_2020}, and molybdenum alloys~\cite{chen_accurate_2017,li_quantum-accurate_2018}.

Other IAP approaches draw on  a range of different ML techniques, such as GPR or KRR~\cite{zeni_building_2018, miwa_molecular_2016, seko_sparse_2014, seko_first-principles_2015, nishiyama_application_2020, pihlajamaki_monte_2020, chapman_multiscale_2020}. Of note is further the (s)GDML approach~\cite{sauceda_molecular_2020, sauceda_molecular_2019, chmiela_sgdml_2019, chmiela_machine_2017, chmiela_towards_2018}, which combines a kernel based ML approach (GDML) with physical symmetries (sGDML) to construct efficient and accurate IAPs. 

Finally, another procedure for constructing ML-IAPs is based on optimizing classical IAPs. This is done in Ref.~\cite{narayanan_machine_2017}, where a genetic algorithm is utilized to fit a bond order potential to investigate metal-organic structures.

\subsubsection{Technical Aspects for the Constructing of Potentials}
\label{sec:iap_technical_aspects}
Finally, there are technical aspects when dealing with ML-IAPs. The construction of these IAPs is often a recurring task, since one is often not only interested in a specific system at specific conditions, but in rather large-scale systematic studies. The methodologies outlined above are not limited to specific elements or conditions apart from limitations introduced by DFT itself. Therefore, frameworks exist to simplify the repeated construction of such potentials. These include PANNA~\cite{lot_panna_2020}, ANI-1~\cite{smith_ani-1_2017-1}, TensorMol~\cite{yao_tensormol-01_2018} and Amp~\cite{khorshidi_amp_2016}, which provide frameworks to build IAPs based on NNs. For MTPs, MLIP~\cite{novikov_mlip_2021} provides a software framework incorporating active learning. GAPs can be build using the QUIP code~\cite{noauthor_quip_nodate} and SNAPs can be constructed with the LAMMPS~\cite{plimpton_fast_1995} code.

Two other closely related aspects worth mentioning in this context are active learning and uncertainty quantification. Ref.~\cite{zhang_active_2019} introduces a framework to realize such active learning procedures for IAPs, while Ref.~\cite{wen_uncertainty_2020} introduces potentials based on dropout-uncertainty neural networks (DUNN) which are capable of assessing uncertainties in their predictions. Such estimations are important, because NNs usually do not perform well during extrapolation tasks, yet such situations might occur during automated materials calculations.

\subsection{Learning Electronic Structures}
\label{sec:learning_electronic_structure}
While the preceding sections discussed ML approaches which center on specific properties extracted from DFT data, be it a wide variety of materials properties (Sec.~\ref{sec:property_mappings}) or specifically the energy-force-landscape (Sec.~\ref{sec:iaps}), the emphasis in the following is on using ML techniques to tackle the electronic structure problem directly. 
In these investigations, DFT is not solely a data acquisition tool, but rather the centerpiece of a ML-powered workflow aimed at a better understanding of electronic structures.

The most common mapping deals with the electronic exchange-correlation energy~\cite{liu_improving_2017,lei_design_2019,chen_deepks-kit_2020,margraf_pure_2021,nelson_machine_2019,schmidt_machine_2019,suzuki_machine_2020,yu_machine_2020,bogojeski_quantum_2020,griego_machine_2020,egorova_multifidelity_2020,dick_machine_2020,fujinami_orbital-free_2020,nagai_completing_2020,mezei_noncovalent_2020,lentz_predicting_2020,sun_toward_2019,zhu_artificial_2019,peccati_regression_2019,hollingsworth_can_2018,vegge_identification_2018,messina_dft-driven_2020}. 
As discussed in Sec.~\ref{sec:dft}, the accuracy of the chosen exchange-correlation approximation primarily determines the accuracy of the overall DFT simulation. The exchange-correlation functionals discussed in Sec.~\ref{sec:dft} mainly employ exact conditions for their constructions, but fitting to either experimental or high-level calculations data is also a common practice for the construction of exchange-correlation functionals. These have been in use for a long time (see Ref.~\cite{becke_density-functional_1988}), especially in quantum chemistry. Using ML methods to perform similar fits is a natural consequence. An example of this is Ref.~\cite{schmidt_machine_2019}. Here, the exact exchange-correlation energy is calculated for a system of limited size and then approximated using a NN, where the electronic density serves as the input quantity. By using automatic differentiation, the exchange-correlation potential is obtained as well. The resulting model incorporates non-local effects, the lack thereof being a usual shortcoming of traditional exchange-correlation functionals at a negligible computational overhead compared to exact calculations. A similar approach is taken in Ref.~\cite{dick_machine_2020}, which introduces NeuralXC, a framework for constructing such NN exchange-correlation functionals, or Ref.~\cite{lei_design_2019}, where CNNs are used to estimate the exchange-correlation energy from the density by drawing on the convolutions of the CNN architecture to featurize the electronic density. They recover the exact exchange of the B3LYP functional.
This is especially important given that the B3LYP functional achieves exact exchange by evaluating the Kohn-Sham orbitals rather then the density, thus demonstrating how NNs can extract meaningful information from the electronic density. Improving exchange-correlation energies with ML is not necessarily limited to the construction of ML functionals. In Ref.~\cite{liu_improving_2017}, range-separation parameters for the established long-range corrected Becke-Lee-Yang-Parr (LC-BLYP) functional \cite{lee_development_1988,becke_density-functional_1988,chiba_excited_2006,song_calculations_2010,tawada_long-range-corrected_2004,iikura_long-range_2001} are learned from molecular features.

Instead of directly learning the exchange-correlation functional itself, corrections to these functionals can be learned. This $\Delta-\mathrm{DFT}$ method was already briefly introduced in Sec.~\ref{sec:quantumchemicalinformation} and has been applied in Ref.~\cite{bogojeski_quantum_2020}. Here, differences between DFT and CC predicted total energies were learned using a KRR model. This model is in turn used to enhance DFT calculations towards chemical accuracy at almost no computational overhead, thereby, enabling highly accurate MD simulations. The advantage of methods is that constructing an ML correction rather then a ML functional reduces the training samples drastically. This is also shown in Ref.~\cite{mezei_noncovalent_2020}, where non-covalent interaction corrections to standard DFT functionals are learned at a much smaller cost.

Using ML to treat density functionals is not limited to exchange-correlation functionals. As introduced in Sec.~\ref{sec:electronic_structure}, the accuracy of computationally efficient OF-DFT simulations mostly depends on the approximation chosen for the kinetic energy functional. Naturally, ML can aid in identifying a suitable functional, and this is subject to current research \cite{hollingsworth_can_2018, fujinami_orbital-free_2020, snyder_finding_2012, meyer_machine_2020}. Also see \cite{li_understanding_2016} for an introduction to the topic. To achieve better performance in this task, in Ref.~\cite{hollingsworth_can_2018} exact conditions are used to constrain ML functionals. Ref.~\cite{li_pure_2016} goes a slightly different route by learning a density functional for the entire total energy.

Going beyond density functionals, other studies focus on intermediate electronic quantities. In Ref.~\cite{yeo_pattern_2019}, pattern-learning techniques are used to predict the density of states (DOS) of an alloy system based on chemical features with high accuracy. The DOS is an important quantity in solid-state physics that can be used to discover or assess the properties of new materials in terms of their band energy. Ref.~\cite{ben_mahmoud_learning_2020} utilizes a similar workflow to predict the DOS, but further includes uncertainty estimates. This workflow is mainly applied to silicon structures. The authors further compare the results of deriving the band energy from a predicted DOS with ML models that directly predict the band energy. They find that the former approach outperforms the latter. This makes intuitive sense given that the ML model learns from a larger amount of data with direct access to information about the electronic structure. 

Another logical quantity of interest in this context is the electronic density itself, for which various ML approaches exist \cite{brockherde_bypassing_2017,tsubaki_quantum_2020,eickenberg_solid_2018,schmidt_learning_2018,alred_machine_2018,grisafi_transferable_2019, fabrizio_electron_2019}. Most notably, Ref.~\cite{brockherde_bypassing_2017} uses KRR to learn and predict electronic densities, although these densities are in turn only intermediate quantities. The principle quantity of interest is the total energy. In learning the electronic density, a choice for the representation of this scalar field has to be made. Both real-space representations \cite{alred_machine_2018} as well as linear expansions in terms of basis functions (either as plane-waves \cite{brockherde_bypassing_2017} or spherical harmonics \cite{grisafi_transferable_2019,fabrizio_electron_2019}) are conceivable, and carry different (dis-)advantages in terms of data footprint and transferability.

Similar to the results of Ref.~\cite{ben_mahmoud_learning_2020}, Ref.~\cite{brockherde_bypassing_2017} finds that learning the electronic density from atomic positions and then performing a separate mapping from electronic density to total energy, leads to more accurate results than can be achieved by a direct mapping of atomic positions to total energies. A separate mapping is needed as the KS-DFT energy functional includes terms that are only implicit functionals of the density. Ref.~\cite{grisafi_transferable_2019} performs a similar study, targeting the exchange-correlation energy rather then the total energy, but finds results opposing those given in Ref.~\cite{brockherde_bypassing_2017}, i.e., the exchange-correlation energy is given more accurately for a direct mapping. As current research \cite{lewis_learning_2021} suggests, these findings might be reconciled by the fact that indirect mappings via the density are more beneficial in extrapolative settings, while direct mappings perform better elsewise. 

In order to gain direct access to the total energy, another target quantity is needed. One possible candidate is the local density of states (LDOS), which gives the DOS at any point in space. In Ref.~\cite{chandrasekaran_solving_2019}, NNs are used to predict both the electronic density and the LDOS individually at each grid point. By integrating the LDOS over the real space afterwards, the DOS is obtained.  
We have recently proposed a similar ML workflow \cite{ellis_accelerating_2021} that follows along similar lines. It uses predicted LDOS values to obtain the total energy from the infered DOS and density. 
Predicted densities can also be used to construct a $\Delta-$DFT~\cite{dick_learning_2019} approach. 

However, electronic structure theory is not limited to an energy-based analysis. Refs.~\cite{hegde_machine-learned_2017} and \cite{ferreira_chemical_2020} apply ML methods to Hamiltonians in order to calculate different quantities of interest. Similarly, Ref.~\cite{panosetti_learning_2020} uses GPR to fit potentials on DFT data, that can be used to improve calculations at lower levels of theory (e.g., density functional tight-binding). Finally, ML can also be used to accelerate the numerical treatment of DFT by reducing the numerical overhead of the solution of the KS equations~\cite{ku_machine_2019} or identifying efficient, adaptive basis sets~\cite{schutt_machine_2018}.

\subsection{Other Approaches}
\label{sec:other_approaches}
There are approaches that do not fit in the categories outlined above, either dealing with more technical aspects on combining ML and DFT (see Sec.~\ref{sec:technical_aspects}) or simply mappings and calculations not discussed yet~\cite{jacobsen_--fly_2018,gao_general_2020,samin_physics-based_2020,venkatraman_can_2015,zhang_nmr-ts_2020,schmidt_computational_2019,choudhary_convergence_2019,frey_prediction_2019,rupp_machine_2014} \cite{takasao_machine_2019,sumita_hunting_2018,tejs_vegge_computational_nodate,tian_effects_2021}. 

An example of such an approach is Ref.~\cite{gao_general_2020}, which is concerned with the already discussed prediction of NMR shifts, but follows a different scheme to directly machine-learn chemical shifts from DFT by using DFT calculated properties as descriptors for a NN. Similar approaches are developed in Ref.~\cite{frey_prediction_2019} and Ref.~\cite{schmidt_computational_2019}.

Contrarily, Ref.~\cite{zhang_nmr-ts_2020} uses DFT as part of a loss function by calculating NMR spectra via DFT for molecules generated by a ML algorithm. Differences in DFT calculated and experimentally measured spectra are used to train this molecular generator. A similar approach is employed in Ref.~\cite{sumita_hunting_2018} for finding organic molecules. Generally, models trained on data are used as key components of such search algorithms in the form of an "evaluator" of candidate structures proposed by a genetic algorithm~\cite{takasao_machine_2019,tejs_vegge_computational_nodate}, an evolutionary algorithm~\cite{jacobsen_--fly_2018}, or a Monte Carlo simulation~\cite{samin_physics-based_2020}.

ML techniques are also used to predict "hyperparameters" of ab-initio simulations. These are usually adjusted manually through experience or by converging them with subsequently more costly and accurate calculations. An example is Ref.~\cite{choudhary_convergence_2019}, where a DT was used to predict the number of k-points and the cutoff energy in plane-wave DFT calculations. Similarly, in Ref.~\cite{venkatraman_can_2015} the choice of the exchange-correlation functional is predicted from chemical descriptors. Finally, DFT-based models are also integrated into elaborate statistical mechanics frameworks as in Ref.~\cite{tian_effects_2021}.


\subsection{Technical Aspects}
\label{sec:technical_aspects}
Building complicated workflows involving DFT and ML poses several technical problems that have to be addressed. First and foremost, any data-driven approach is fundamentally reliant on data. In the context of ab-initio simulations, data is both costly to acquire and digitally exchangeable by nature. To that end, it makes intuitive sense to recycle already pre-calculated data and models in terms of databases, a number of which are given in Tab.~\ref{tab:usefulSoftwares}, alongside a short overview  over useful software frameworks for ML-DFT approaches.
For example, JARVIS includes properties not only on around 40,000 materials, but also ready-to-use ML models trained on DFT data. The database is constantly expanded and, thereby, facilitates the construction of new workflows. Similarly, acquiring new training data is facilitated with integrated workflows~\cite{nandy_machine_2019}.

Another important aspect concerns the choices with respect to the ML methodologies. A wealth of ML techniques can be applied to DFT calculations and data. Likewise, there is a sizable amount of procedures to encode information in terms of descriptors. Investigations~\cite{christopher_j_bartel_critical_2020,agarwal_understanding_2020,hou_comparison_2018}
as to which of these techniques hold advantages over others are crucial. 
In Ref.~\cite{hou_comparison_2018} different types of neural networks (CNN, DNN, single-layer NN) have been assessed in terms of predicting molecular properties, with a shallow DNN performing best. Similarly, in Ref.~\cite{christopher_j_bartel_critical_2020}, different ways to encode compositional information for a GB model were compared. None of the investigated methods was able to produce models capable of assessing the stability of new materials reliably.

Indeed, the development of new descriptors and techniques to represent chemical information for specialized use cases is subject of ongoing research~\cite{collins_effective_2020,gusarov_development_2020,dickel_neural_2020,laghuvarapu_band_2020,sahu_toward_2018,choudhary_machine_2018,li_correlation_2018,jalem_general_2018,deimel_active_2020,townsend_representation_2020}. 
For example, Ref.~\cite{townsend_representation_2020} introduces persistent image descriptors (PI), based on concepts from applied mathematics, to represent chemical structures in the search for functional groups capable of capturing $\mathrm{CO}_2$. Different ML models were trained on data encoded by these descriptors and compared to other commonly used descriptors.\\
Common descriptors are  SOAP~\cite{bartok_representing_2013}, the Coulomb matrix~\cite{rupp_fast_2012}, BoB~\cite{hansen_machine_2015}, FCHL~\cite{faber_alchemical_2017}, and ACE~\cite{drautz_atomic_2019, lysogorskiy_performant_2021}. This list is not comprehensive, and a number of ML workflows use customized descriptors or simply easily accessible chemical information to represent materials. The utility of these descriptors is demonstrated in the aforementioned publications. Ref.~\cite{rupp_fast_2012, faber_alchemical_2017, hansen_machine_2015} use their respective descriptors and KRR to learn and predict atomization energies. On the other hand, Ref.~\cite{bartok_representing_2013} uses SOAP descriptors to construct and compare GAPs, with Ref.~\cite{drautz_atomic_2019} building ACE based IAPs as well.

Lastly, transferability and uncertainty quantification have to be taken into account in order to apply combined ML-DFT workflows in future applications. Correctly treating such aspects reduces the need to retrain models. As Ref.~\cite{tamura_machine_2019} suggests, the transferability of ML models can be utilized with certain limitations, as demonstrated for moderate volume changes in the liquid phase. Uncertainty quantification, as investigated in \cite{peterson_addressing_2017}, deals with assessing whether a particular prediction can be trusted or additional data is needed to improve the model.

\begin{table*}[htp]
    \centering  
    \caption{Selection of useful software frameworks for ML-DFT.}\label{tab:usefulSoftwares}		
    \begin{tabular}{ll}
    	\toprule
    	\textbf{Task} &  \textbf{Software}  \ \\\midrule
    	DFT calculations (periodic systems) & VASP \cite{kresse_efficient_1996, kresse_efficiency_1996, kresse_ab_1993}, QuantumESPRESSO \cite{giannozzi_quantum_2009, giannozzi_q_2020, giannozzi_advanced_2017}, WIEN2k \cite{blaha_wien2k_2001, blaha_wien2k_2020}, CASTEP \cite{clark_first_2005} \\
    	DFT calculations (isolated systems) & Gaussian \cite{frisch_gaussian_2016}, GAMESS \cite{schmidt_general_1993, gordon_advances_2005}, ORCA \cite{neese_orca_2012, neese_software_2018}, MOLPRO \cite{werner_molpro_2012, werner_molpro_2019, knowles_new_1984, werner_second_1985, knowles_efficient_1985, werner_efficient_1988, knowles_efficient_1988, amos_open_1991, knowles_restricted_1991, knowles_internally_1992, knowles_coupled_1993, knowles_determinant_1989} \\ 
    	Neural Network construction & PyTorch \cite{paszke_pytorch_2019, noauthor_pytorch_nodate}, TensorFlow \cite{martin_abadi_tensorflow_2015}, Keras \cite{chollet_keras_2015}, Flux.jl \cite{flux1, flux2} \\
    	General ML library & Shogun \cite{sonnenburg2010shogun}, Scikit-Learn \cite{pedregosa_scikit-learn_2011}, Weka \cite{witten2002data} \\ 
    	Gaussian Process Regression & GPytorch \cite{gardner_gpytorch_2018} \\ \midrule 
    	Constructing NN-IAPs & PANNA~\cite{lot_panna_2020}, ANI-1~\cite{smith_ani-1_2017-1}, TensorMol~\cite{yao_tensormol-01_2018}, Amp~\cite{khorshidi_amp_2016}\\ 
    	Constructing other IAPs & MLIP~\cite{novikov_mlip_2021}, QUIP code~\cite{noauthor_quip_nodate}, LAMMPS~\cite{plimpton_fast_1995}  \ \\ 
    	ML-exchange-correlation & NeuralXC~\cite{dick_machine_2020} \\ \midrule 
    	Toolchaining & ASE~\cite{hjorth_larsen_atomic_2017}, AiiDA~\cite{aiida1,aiida2}, DeepChem~\cite{DeepChem} \\ 
    	Databases & MoleculeNet~\cite{wu_moleculenet_nodate}, OQMD~\cite{saal_materials_2013}, ANI-1~\cite{smith_ani-1_2017}, PubChemQC~\cite{nakata_pubchemqc_2017}, NOMAD~\cite{draxl_nomad_2018}, \\ ~ & JARVIS~\cite{choudhary_joint_2020}, OC20 \cite{chanussot_open_2020} \ \\ \bottomrule
    \end{tabular}
\end{table*}			

\section{Discussion}
\label{sec:discussion}

\subsection{Citation Analysis}
\label{sec:citation_analysis}
\begin{figure*}[htp]
    \centering
    \includegraphics[width=1.9\columnwidth]{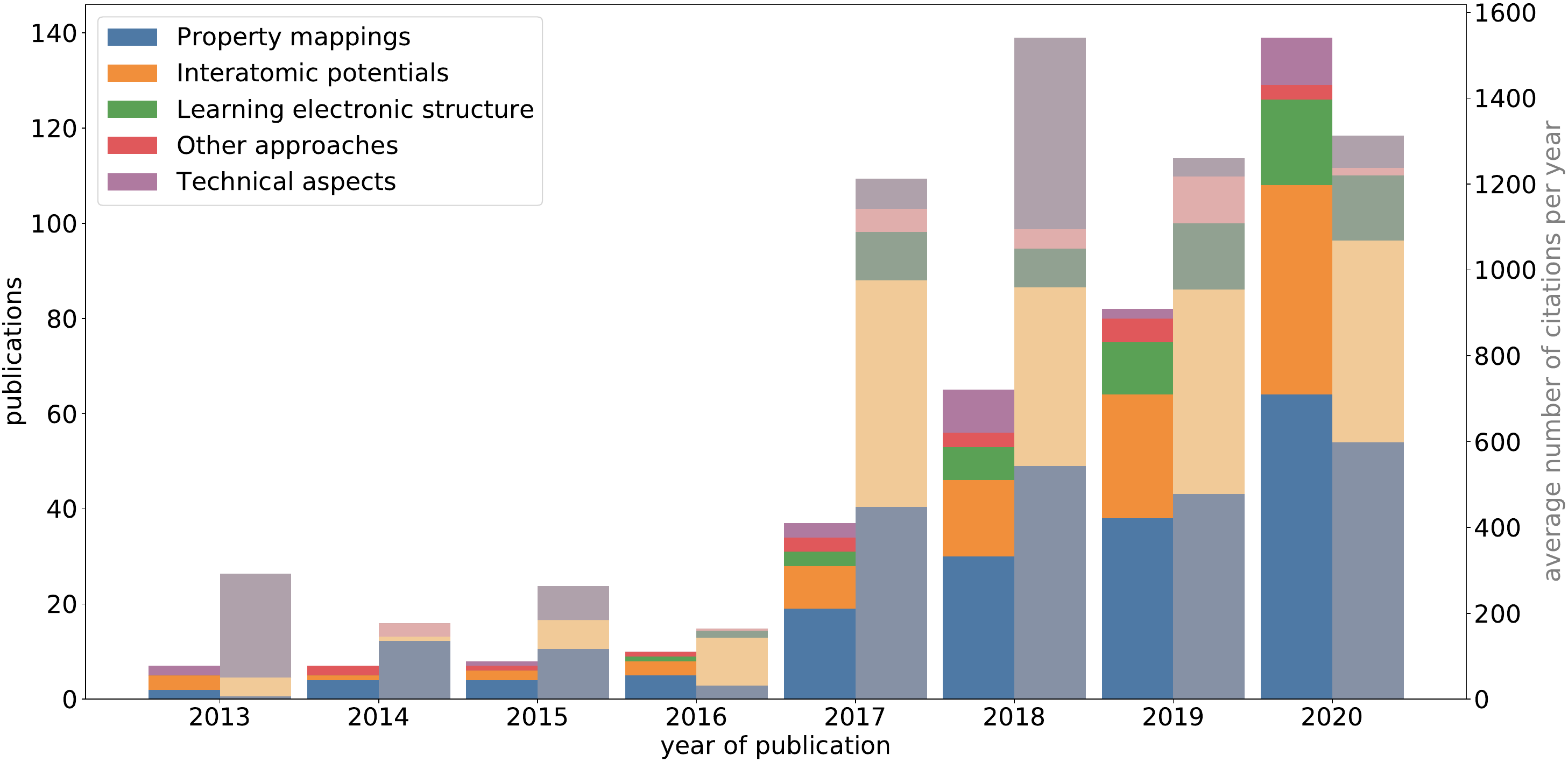}
    \caption{Number of publications and average number of citations per year by year of publication and category.}
    \label{fig:all_categories_citations}
\end{figure*}

\begin{figure}[ht]
    \centering
    \includegraphics[width=0.9\columnwidth]{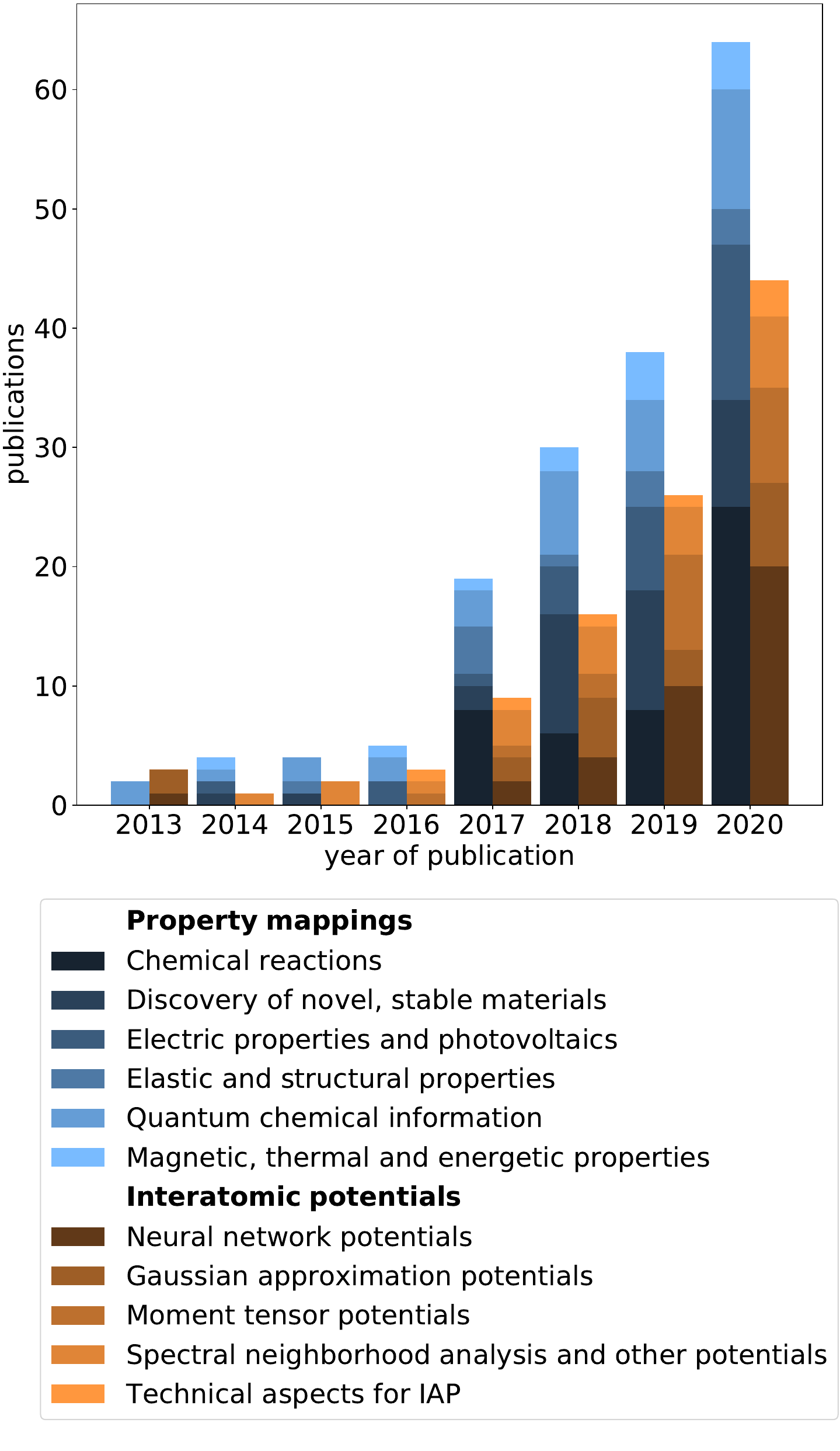}
    \caption{Number of publications by year for property mapping and interatomic potential related publications.}
    \label{fig:c1c2_publications}
\end{figure}

We conclude this review by assessing general trends in the field in terms of a citation analysis based on the 
370 research articles collected. The first part of this collection was gathered at the end of 2020 using Web of Science \cite{noauthor_web_1997}. Grouping of these articles was performed manually. Citation numbers where retrieved via Google Scholar \cite{noauthor_google_2004} by an automated python script. Both the script and the database of research articles are provided alongside this publication \cite{fiedler_dataset_2021}. 

The results of this citation analysis are shown in Fig.~\ref{fig:all_categories_citations}. We note that we use the average number of citations as a measure in Fig.~\ref{fig:all_categories_citations}. It denotes how often a research article published in a particular year was \textit{cited per year} on average in subsequent years. In this manner, articles published in later years can be judged on the same basis.

It is apparent from Fig.~\ref{fig:all_categories_citations} that the number of publications increases rapidly, highlighting the growing importance of this research area. The number of publications has more then tripled in the years between 2017 and 2020. Interestingly, this growth in number of publications is not necessarily reflected in the number of citations.
The average number of citations is consistent for works published in 2017--2020, despite the total number of articles growing rapidly in these years. A possible explanation lies in the relatively young age of this research area. Recently published research articles are consequently very relevant for more recent publications. 
A clear illustration of this fact is apparent in the set of publications in 2013, where only two articles, Ref.~\cite{bartok_representing_2013} and \cite{saal_materials_2013}, have accumulated a large number of citations.

Fig.~\ref{fig:all_categories_citations} further illustrates how most research is directed towards property mappings and interatomic potentials. This makes intuitive sense, given that DFT calculations often underpin large-scale simulations that are based on data-driven methodologies. Yet, also a growing number of publications is concerned with applying ML to the electronic structure problem itself.

\begin{figure}[ht]
    \centering
    \includegraphics[width=0.9\columnwidth]{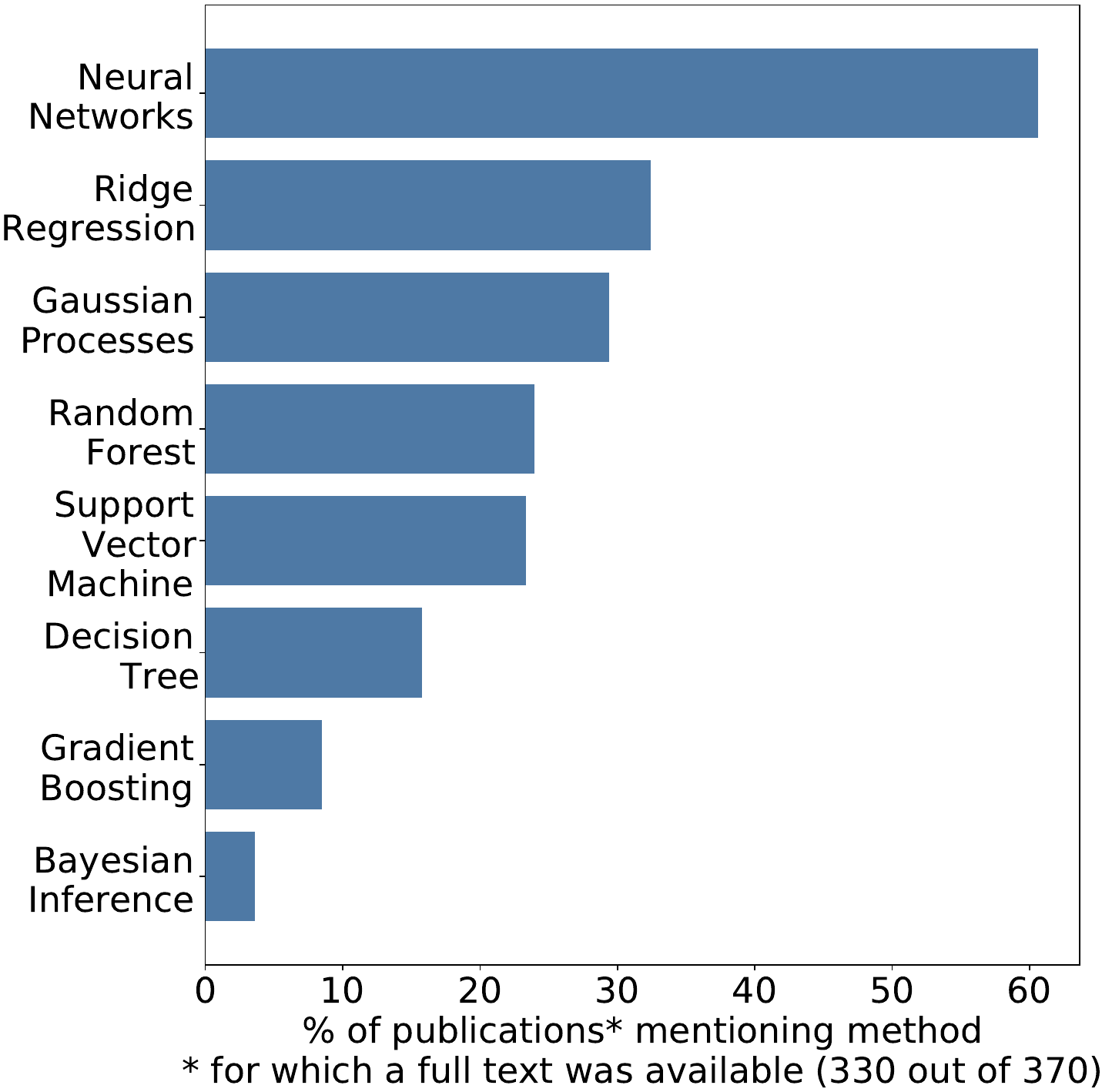}
    \caption{Mentions of ML methods throughout the reviewed articles.}
    \label{fig:ml_methods}
\end{figure}

We perform the same analysis on the two largest categories in Fig.~\ref{fig:all_categories_citations} separately. The results are illustrated in Fig.~\ref{fig:c1c2_publications}. Except for an increasing interest in applications concerning electrical properties and photovoltaics, another dominating trend cannot be identified for property mapping applications. Likewise, it can be seen that while in 2020 nearly half of all publications using IAPs employed NNs, the relative number of publications for the respective types of IAPs are relatively evenly distributed throughout the years. The large number of NN related publications is further illustrated in Fig.~\ref{fig:ml_methods}, which analyzes the number of publications mentioning certain ML methods. A prevalence for NNs is apparent, with Ridge Regression and GPR being the second and third-most used methods, respectively.

\subsection{Conclusions and Outlook}
An extensive review over research combining ML and DFT was performed. By compiling and analyzing over 300 peer-reviewed research articles, we were able to identify five principal categories of research, presented in Sec.~\ref{sec:review_section}. 
Most current efforts are focused on using data-driven methodologies to assist conventional DFT studies of materials, either by directly working with application-specific quantities (Sec.~\ref{sec:property_mappings}) or by providing ML-IAPs (Sec.~\ref{sec:iaps}) to facilitate dynamical calculations. The addressed range of topics is vast and follows pressing scientific problems, such as novel materials discovery, reducing the human carbon footprint, or enabling novel energy solutions. A number of publications (see Secs.~\ref{sec:iap_technical_aspects} and \ref{sec:technical_aspects}) is concerned with technical aspects of enabling ML-DFT workflows by providing fundamental software solutions. Of particular importance are databases, which are the backbone for any data-driven approach. Another, still underrated research area focuses directly on addressing the shortcomings of DFT as an electronic structure method. These efforts have been discussed in Sec.~\ref{sec:learning_electronic_structure}. Drawing on this analysis, a general ML-DFT workflow can be identified, as is shown in Fig.~\ref{fig:ml_workflow}, representing the vast majority of ML-DFT approaches, with only a small number of very specialized methods not being included. Both the variety as well as similarities in different methods can be identified in Fig.~\ref{fig:ml_workflow}.
In general, the research interest in ML-DFT methods is growing fast, as shown in Sec.~\ref{sec:citation_analysis}, and methods addressing the electronic structure problem directly are gaining traction. Considering the rising number of publication, this research field is expected to grow further. Future investigations may address topics such as uncertainty quantification and active learning approaches. Thereby, they may broaden the utility of ML in computational material science and computational chemistry.

With ever improving ML-DFT models, new applications that were previously unattainable will become possible, such as large-scale automated materials discovery, multi-scale modelling of materials, and digital twins of complex systems. 

\section*{Acknowledgments}
This work was funded by the Center for Advanced Systems Understanding (CASUS) which is financed by the German Federal Ministry of Education and Research (BMBF) and by the Saxon Ministry for Science, Art, and Tourism (SMWK) with tax funds on the basis of the budget approved by the Saxon State Parliament. We thank Aidan Thompson and Sebastian Schwalbe for useful communications.
\clearpage



%

\end{document}